# From Finite Enumeration to Universal Proof: Ring-Theoretic Foundations for PQC Hardware Masking Verification

Ray Iskander[1], Khaled Kirah[2,*]

[1]Verdict Security
[2]Faculty of Engineering, Ain Shams University, Cairo, Egypt

**Abstract**

Formal verification of masking in post-quantum cryptographic (PQC) hardware relies on SMT solvers over finite domains. Our prior work established structural dependency analysis at scale [1] and quantified the security margin of partial NTT masking [2]. QANARY, our structural dependency analysis framework, verified 1.17 million cells across 30 modules of the Adams Bridge ML-DSA/ML-KEM accelerator [3, 4], but its core soundness result (Theorem 3.9.1) was machine-checked only at $q = 5$ via $2^{25}$ Boolean wire functions. This left portability to ML-KEM ($q = 3{,}329$, FIPS 203 [5]) and ML-DSA ($q = 8{,}380{,}417$, FIPS 204 [6]) as an open gap. NIST IR 8547 [7] (March 2025) motivates closing such gaps. We present the first machine-checked universal proof of the $r$-free sub-theorem of Theorem 3.9.1: for every $q > 0$, every wire function, and every pair of secrets, value-independence implies identical marginal distributions. The proof, in Lean 4 [8] with Mathlib [9], requires five lines versus $2^{25}$ finite evaluations. It is sorry-free, reducing the trusted base from {Z3 [10], CVC5 [11], Python} to the Lean 4 kernel. We provide nine theorems (T1–T6, T1', T3') covering reparametrization, bijectivity, overflow bounds, RNG bias, and a universal non-tightness counterexample for all $q \geq 2$. The results establish commutative ring axioms of $\mathbb{Z}/q\mathbb{Z}$ as the natural abstraction layer for arithmetic masking verification.

## 1. Introduction

The finalization of ML-KEM (FIPS 203) and ML-DSA (FIPS 204) in August 2024 marks the beginning of a large-scale migration to post-quantum cryptography. NIST IR 8547, published in March 2025, establishes the transition timeline and recommends formal methods for validating PQC implementations. For hardware accelerators implementing these standards, masked arithmetic over $\mathbb{Z}_q$ , particularly within Number Theoretic Transform (NTT) butterflies, is the primary defense against power side-channel attacks.

This paper is the third in a series addressing PQC hardware security from complementary angles: scalable verification tooling (Paper 1) [1], attack-surface quantification (Paper 2) [2], universal formal foundations (this paper), and probabilistic completeness (forthcoming). Each paper in the series is self-contained; the present paper can be read without prior familiarity with Papers 1–2. Readers seeking the practical motivation for the universal proof, namely, why SMT-verified-at-$q = 5$ was insufficient, will find it developed in Paper 2. In paper [1], we introduced QANARY, a structural dependency analysis framework that scaled to 1.17 million cells across 30 modules and identified structural vulnerabilities independently confirmed in Ref. [12] (see [1] for details). In paper [2], we demonstrated that partial NTT masking in the Adams Bridge accelerator leaves security margins 25–29 bits smaller than claimed.
QANARY's soundness rests on Theorem 3.9.1: if a wire function is *value-independent* of the secret, then its marginal distribution is constant, implying zero mutual information with the secret. In [2], this theorem was machine-checked by Z3 and CVC5 at modulus $q = 5$ by exhaustively enumerating all $2^{25}$ Boolean wire functions over $\mathbb{Z}_5 \times \mathbb{Z}_5$. This finite-domain verification, while providing strong evidence, cannot structurally guarantee correctness at the production moduli $q = 3{,}329$ (ML-KEM) or $q = 8{,}380{,}417$ (ML-DSA). SMT solvers enumerate; they cannot perform structural induction.

*Correspondence Author: khaled.kirah@eng.asu.edu.eg
Ray Iskander: ray@verdictsecurity.com

This paper closes that gap. We contribute:

1. **A universal Lean 4 proof of the $r$-free sub-theorem of Theorem 3.9.1**, valid for all $q > 0$, all wire functions $w: \mathbb{Z}_q \times \mathbb{Z}_q \to \beta$ (for arbitrary output type $\beta$), and all secrets $x, x'$. The proof requires five lines of tactic script (Section 4). By the law of total probability, establishing this $r$-free algebraic property is the exact prerequisite for full probabilistic security.
2. **Universal proofs of six supporting theorems (T2–T6):** reparametrization round-trips for Boolean and arithmetic masking, reparametrization bijectivity, overflow bounds bridging to bit-vector implementations, RNG bias characterization, and a universal non-tightness counterexample showing that the converse of Theorem 3.9.1 fails for every $q \geq 2$ (Section 5).
3. **A methodological insight:** the five-line proof follows from commutative ring axioms alone (sub_add_cancel in Mathlib's CommRing instance for $\mathbb{Z}/q\mathbb{Z}$). The secret-mask reparametrization is a ring identity, not a computational procedure. This reveals that ring theory, not bit-vector SAT, is the natural abstraction layer for arithmetic masking verification (Section 6).

We emphasize that the simplicity of the proof is the point, not a weakness. The Z3/CVC5 approach required $2^{25}$ evaluations not because the property is complex, but because bit-vector SAT obscures the algebraic structure that makes it straightforward. The five-line proof is evidence that the SADC framework has a natural algebraic foundation, and that prior finite-domain approaches were working in the wrong formalism. For practitioners, this means that QANARY's soundness guarantee now covers every NIST PQC parameter set, present and future, without re-verification.

The entire proof suite is sorry-free: 1,739 Lean 4 build tasks (as reported by lake build), zero sorry, zero errors, on Lean 4.30.0-rc1 with Mathlib pinned to commit 322515540d7f. The main theorem and supporting lemmas T1'–T4, T5 universal bounds, and T6 are kernel-verified; T5's ML-KEM instance uses native compilation (see Section 5.4). The artifact is publicly available at https://github.com/rayiskander2406/qanary-universal-masking-proofs-arXiv-2604.18717 and archived on Zenodo (see Code and Data Availability).

**Paper organization.** Section 2 reviews the QANARY verification hierarchy and the finite-domain gap. Section 3 surveys related work on formal masking verification. Section 4 presents the main theorem. Section 5 presents the supporting proof suite. Section 6 discusses implications for certification, portability, and tool trust. Section 7 addresses limitations and future work. Section 8 concludes.

# 2. Background

## 2.1. Arithmetic Masking in NTT Hardware

First-order arithmetic masking splits a secret $x \in \mathbb{Z}_q$ into two shares $s_0, s_1 \in \mathbb{Z}_q$ satisfying $s_0 + s_1 \equiv x \pmod{q}$. An adversary observing a single wire (the standard probing model [13]) should learn nothing about $x$.

**Definition 1 (Wire Function).** A *wire function* is a map $w: \mathbb{Z}_q \times \mathbb{Z}_q \to \beta$ where $\beta$ is an arbitrary output type. The inputs are shares $(s_0, s_1)$.

**Definition 2 (Arithmetic Reparametrization).** Given secret $x$ and mask $s_1$, the *arithmetic reparametrization* computes $s_0 = x - s_1$ in $\mathbb{Z}_q$. This re-expresses $w(s_0, s_1)$ as $w(x - s_1, s_1)$, a function of $(x, s_1)$.

**Definition 3 (Value-Independence).** A wire function $w$ is *value-independent* of $x$ if

$$\forall\, s_1, x, x' \in \mathbb{Z}_q\colon\ w(x - s_1,\, s_1) = w(x' - s_1,\, s_1).$$

When $s_1$ is drawn uniformly from $\mathbb{Z}_q$, value-independence implies that the conditional distribution of $w$ given $x$ does not depend on $x$, yielding $I(x; w) = 0$. This is the distributional security guarantee that QANARY's structural dependency analysis (SADC) checks.

## 2.2. The QANARY Verification Hierarchy

QANARY organizes verification into four stages of increasing precision:

1. **D0/D1 queries:** for each wire, determine which shares influence its value. A wire depending only on $s_0$ or only on $s_1$ is trivially secure.
2. **Fresh masking (FM):** identify wires protected by fresh randomness, which masks the dependence on the secret.
3. **Boolean SADC:** for Boolean-masked domains, check value-independence via the XOR reparametrization.
4. **Arithmetic SADC:** for arithmetic-masked domains (NTT butterflies, Barrett reduction), check value-independence via the modular subtraction reparametrization (Definition 2).

Theorem 3.9.1 is the soundness guarantee for stage 4: any wire that SADC labels SECURE (i.e., value-independent) provably leaks no information about the secret.

## 2.3. The Finite-Domain Gap

In Paper 2, Theorem 3.9.1 was verified by encoding it in the quantifier-free bit-vector theory (QF_BV) and checking with Z3 and CVC5 at $q = 5$. At this modulus, the wire function space $w: \mathbb{Z}_5 \times \mathbb{Z}_5 \to \{0,1\}$ has $2^{25}$ elements (one Boolean output per input pair). Both solvers confirmed the theorem with zero disagreements across 363 dual-solver checks.

This verification is strong evidence but has three structural limitations:

1. **No induction.** SMT solvers cannot perform structural induction over $q$. A proof at $q = 5$ is logically independent of a proof at $q = 3{,}329$.
2. **Exponential scaling.** At $q = 3{,}329$, the wire space has $2^{3329^2} = 2^{11{,}082{,}241}$ elements, far beyond feasible enumeration.
3. **Trusted base.** Correctness depends on the absence of bugs in Z3, CVC5, and the Python encoding that generates the SMT-LIB2 queries.

Paper 2 explicitly listed this finite-domain ceiling as the one remaining open methodological gap (Section 6, Limitation (ii)). This paper closes it.

## 2.4. Interactive Theorem Proving with Lean 4

Lean 4 is a dependently typed programming language and interactive theorem prover (ITP). Mathlib, its mathematical library, provides the type ZMod q: the quotient ring $\mathbb{Z}/q\mathbb{Z}$ with a CommRing instance. Arithmetic in ZMod q is modular by construction, there are no bit-vector overflow concerns.

A Lean 4 proof is kernel-verified: the Lean kernel independently checks every proof term, providing a minimal trusted base. A proof marked sorry-free contains no unverified stubs (sorry or admit).

# 3. Related Work

Table 1 summarizes the landscape. We distinguish three axes: the masking domain (Boolean vs. arithmetic), whether the result is universal over parameters, and whether the formalization targets NTT-specific hardware.

**Table 1.** Comparison of formal masking verification approaches. "Universal" means the result holds for all parameter values (modulus, bit-width), not just specific instances. "NTT-specific" means the formalization addresses arithmetic operations within NTT butterflies.

| Tool / Work | Venue | Proof System | Masking | Universal | NTT-Specific | Kernel-Verified |
|---|---|---|---|---|---|---|
| maskVerif [14] | EUROCRYPT 2015 | EasyCrypt | Boolean | No | No | Yes (EasyCrypt) |
| REBECCA [15] | EUROCRYPT 2018 | — | Boolean | No | No | No |
| SILVER [16] | ASIACRYPT 2020 | Isabelle/HOL | Boolean | Partial (circuit, not field)[a] | No | Yes (Isabelle) |
| Coco [17] | USENIX 2021 | Isabelle/HOL | Boolean | No | No | Yes (Isabelle) |
| Ref. [18] | ACNS 2023 | Isabelle/HOL | Arithmetic | Partial (circuit, not field)[a] | No | Yes (Isabelle) |
| QANARY [1] | — | Z3 / CVC5 | Arithmetic | No[b] | Yes | No (SMT) |

| Tool / Work | Venue | Proof System | Masking | Universal | NTT-Specific | Kernel-Verified |
|---|---|---|---|---|---|---|
| **This work** | — | **Lean 4 / Mathlib** | **Arithmetic** | **Yes** | **Yes** | **Yes (Lean)** |

[a]Isabelle/HOL soundness proofs are universal over circuit structure but not over the prime field modulus.
[b]Verified at $q = 5$ only.

While powerful underlying ITPs like Isabelle/HOL and EasyCrypt inherently support arbitrary algebraic structures, the formal verification *tools* built upon them for hardware masking have historically targeted Boolean domains to protect symmetric primitives (AES, Keccak).

**Boolean masking verification.** maskVerif [14] was introduced with EasyCrypt proofs for software implementations, and subsequently formalized Strong Non-Interference (SNI) as a compositional security notion for masked gadgets [19]. REBECCA [15], was developed providing netlist-level verification with glitch awareness. SILVER [16] has ROBDD-based exact verification and Isabelle/HOL soundness proofs. Gigerl et al. extended this line to co-verification of masked software on CPUs (Coco [17]) and, most recently, to the formal verification of arithmetic masking in hardware and software [18], the closest prior work to ours, but still circuit-oriented rather than universal over the prime field modulus. These works do not address machine-checked universal algebraic properties of masked arithmetic over arbitrary $\mathbb{Z}_q$, which is the core operation in NTT-based PQC hardware.

**Domain-oriented masking and mask conversion.** Related hardware primitives include domain-oriented masking (DOM) [20], which provides compact pipelined gadgets with arbitrary protection order, and sound methods for switching between Boolean and arithmetic masking [21]. Our contribution is orthogonal to these: we establish universal algebraic properties of masked arithmetic in the probing model [13], not new gadget constructions or conversion algorithms.

**Gap.** No prior work provides a machine-checked proof of arithmetic masking security that is simultaneously (a) universal over the prime field modulus, (b) specific to NTT hardware operations, and (c) kernel-verified by an ITP. This is the gap we close.

## 4. The Main Theorem

**Theorem 4.1 (Value-Independence Implies Constant Marginal; Universal).** Let $q > 0$. Let $\beta$ be any type with decidable equality. Let $w\colon \mathbb{Z}_q \times \mathbb{Z}_q \to \beta$ be a wire function. Define the *marginal histogram*

$$h(x, v) \ := \ |\{s_1 \in \mathbb{Z}_q \colon w(x - s_1,\, s_1) = v\}|.$$

If $w$ is value-independent (Definition 3), then for all $x, x' \in \mathbb{Z}_q$ and all $v \in \beta$:

$$h(x, v) = h(x', v).$$

**Theorem 4.1' (T1': General Output Type).** The same result holds for wire functions with arbitrary output type $\beta$ with decidable equality. The proof is identical: value-independence makes the filter predicate independent of $x$ for any $\beta$. In the artifact, value_independence_implies_constant_histogram is stated with {β : Type*} [DecidableEq β], with T1 as the Bool specialization.

**Observation 4.2 (Informal).** Under value-independence, $P(w = v \mid x) = h(x, v)/q$ is constant in $x$, so $I(x; w) = 0$. (This information-theoretic conclusion follows by a standard argument from the constant histogram; it is not formalized in the artifact. See Limitation (ii).)

## 4.1. The Five-Line Proof

The following is the complete Lean 4 proof of Theorem 4.1, reproduced verbatim from the artifact:

```
theorem value_independence_implies_constant_histogram
    {q : ℕ} [NeZero q] {β : Type*} [DecidableEq β]
    (w : WireFunction q β) (hw : ValueIndependent w)
    (x x' : ZMod q) (v : β) :
    marginalHistogram w x v
      = marginalHistogram w x' v := by
  unfold marginalHistogram
  congr 1
  ext s₁
  simp only [mem_filter, mem_univ, true_and]
  exact ⟨fun h => by rw [← hw s₁ x x']; exact h,
       fun h => by rw [hw s₁ x x']; exact h⟩
```

## 4.2. Why Five Lines Suffice

The proof proceeds in four logical steps:

1. **Unfold the definition.** The marginal histogram counts elements of a filtered finite set: those $s_1 \in \mathbb{Z}_q$ for which $w(x - s_1, s_1) = v$.
2. **Reduce to set equality.** Two finite sets with the same elements have the same cardinality (congr 1 reduces the cardinality goal to a set-extensionality goal).
3. **Pointwise equivalence.** For each $s_1$, the filter predicate $w(x - s_1, s_1) = v$ holds if and only if $w(x' - s_1, s_1) = v$. This is immediate from value-independence: $w(x - s_1, s_1) = w(x' - s_1, s_1)$.
4. **Close the iff.** The two directions of the biconditional each follow by a single rewrite using the value-independence hypothesis.

The Z3/CVC5 proof at $q = 5$ required checking $2^{25} = 33{,}554{,}432$ Boolean wire functions because SMT solvers must instantiate universal quantifiers. Lean 4's dependent type theory handles the universal quantification natively: the proof works for *any* $q$, *any* $w$, *any* $\beta$, simultaneously.

The brevity is not an artifact of library complexity absorbing the difficulty. The proof uses only Finset.mem_filter, Finset.mem_univ, and the hypothesis hw. No deep Mathlib theorems are invoked. The result is short because the property is a structural consequence of the definition, value-independence is the statement that the filter predicate does not depend on $x$.

## 5. The Supporting Proof Suite

Figure 1 shows the logical dependencies among the theorems. All nine theorems are sorry-free. T1, T1’, T2, T3, T3’, T4, T5 universal bounds, and T6 are kernel-verified. T5’s ML-KEM instance is verified by native compilation (native_decide); see Section 5.4. Table 2 summarizes each theorem’s scope, key proof technique, and the parameter coverage it replaces from Paper 2’s finite-domain baseline.

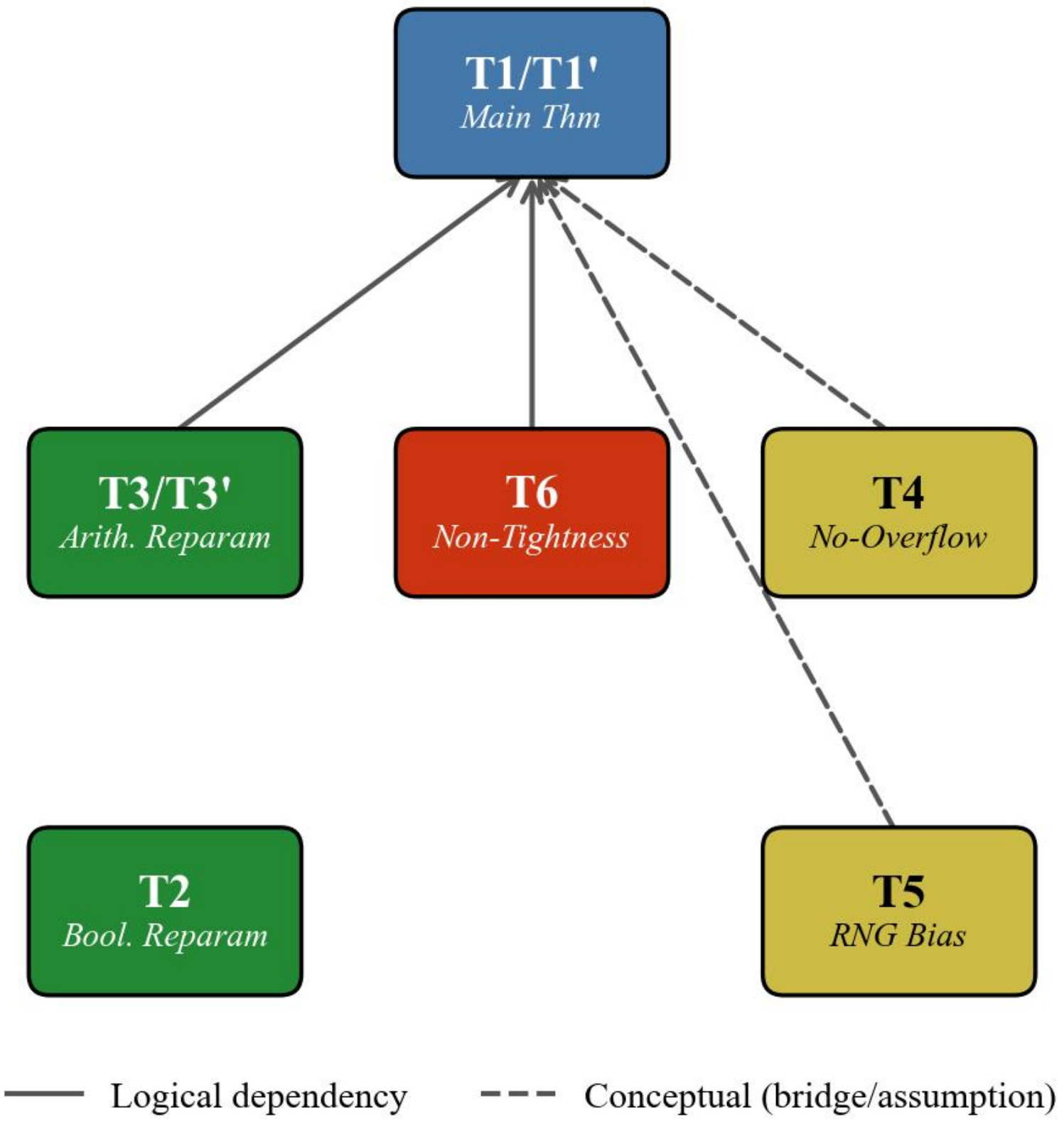

*Figure 1. Proof suite dependency structure. Solid arrows: logical dependency. Dashed arrows: conceptual relationship (T4 bridges to hardware encoding; T5 validates the uniformity assumption). T6 establishes non-tightness of T1.*

**Table 2.** Complete proof suite summary. All theorems are sorry-free and kernel-verified in Lean 4.30.0-rc1 with Mathlib 322515540d7f.

| ID | Theorem | Domain | Key Technique | Upgrade from Paper 2 |
|---|---|---|---|---|
| T1 | Value-independence implies constant marginal (Bool) | All $q > 0$ | congr + ext + rewrite | $q = 5, 2^{25}$ wires |
| T1’ | Value-independence implies constant histogram (general $\beta$) | All $q > 0$ | Same as T1 | New |

| ID | Theorem | Domain | Key Technique | Upgrade from Paper 2 |
|---|---|---|---|---|
| T2 | Boolean reparametrization round-trip | All $k$ | XOR assoc + self-inverse | $k = 24$ |
| T3 | Arithmetic reparametrization round-trip | All $q > 0$ | sub_add_cancel | $q \in \{3329,838041$ |
| T3’ | Reparametrization bijectivity | All $q > 0$ | Injective + surjective | New |
| T4 | No-overflow bounds | All $q > 0$ | omega | 3 configurations |
| T5 | ML-KEM bias ratio | $q = 3329$ | native_decide | Python enumeration |
| T5 (bounds) | Universal RNG bias bounds | All $q > 0$ | card_le_card_of_injOn | New |
| T6 | Non-tightness counterexample | All $q \geq 2$ | card_bij + one_ne_zero | $q = 5$ only |

**Reproduction.** Clone the repository. Ensure the lean-toolchain file reads leanprover/lean4:v4.30.0-rc1. Run lake build. Expected output: 1,739 tasks completed, 0 errors. All theorem names in this paper correspond to identifiers in QanaryUniversal/*.lean. The lakefile.lean pins Mathlib to commit 322515540d7f; no other dependencies are required. We intend to submit the artifact for venue artifact evaluation.

## 5.1. T2: Boolean Reparametrization Round-Trip

**Theorem 5.1 (T2).** For all $k \in \mathbb{N}$ and all $x, s_1 \in \{0,1\}^k$: $(x \oplus s_1) \oplus s_1 = x$.

```
theorem bool_reparam_round_trip
    {k : ℕ} (x s₁ : BitVec k) :
    boolReparam (boolReparam x s₁) s₁ = x := by
  simp [boolReparam, BitVec.xor_assoc,
      BitVec.xor_self, BitVec.xor_zero]
```

This upgrades the Paper 2 version from $k = 24$ to arbitrary $k$. The proof follows from XOR associativity, self-inverse ($a \oplus a = 0$), and identity ($a \oplus 0 = a$).

## 5.2. T3: Arithmetic Reparametrization Round-Trip

**Theorem 5.2 (T3).** For all $q > 0$ and all $x, s_1 \in \mathbb{Z}_q$: $(x - s_1) + s_1 = x$.

```
theorem arith_reparam_round_trip
    {q : ℕ} [NeZero q] (x s₁ : ZMod q) :
    arithReparam q x s₁ + s₁ = x := by
  simp [arithReparam, sub_add_cancel]
```

This is the key algebraic step. In Paper 2, the analogous result was encoded in QF_BV as $\text{URem}(\text{URem}(x - s_1 + q,\ q) + s_1,\ q) = x$ with explicit overflow checks, verified only for $q \in \{3{,}329,\ 8{,}380{,}417\}$ at bit-width $w = 24$. In Lean 4, $\mathbb{Z}_q$ is the quotient ring $\mathbb{Z}/q\mathbb{Z}$; the identity sub_add_cancel is a CommRing axiom. The overflow concern evaporates, it was an artifact of the bit-vector encoding.

**Theorem 5.3 (T3': Bijectivity).** For all $q > 0$ and each fixed $s_1 \in \mathbb{Z}_q$, the map $x \mapsto x - s_1$ is a bijection on $\mathbb{Z}_q$.

```
theorem arith_reparam_bijective
    {q : ℕ} [NeZero q] (s₁ : ZMod q) :
    Function.Bijective
      (fun x => arithReparam q x s₁) := by
  constructor
  · intro x x' h
    simp [arithReparam] at h; exact h
  · intro s₀
    exact ⟨s₀ + s₁, by simp [arithReparam]⟩
```

Bijectivity ensures that as $x$ varies over $\mathbb{Z}_q$, so does $s_0 = x - s_1$, no collisions occur. This is a prerequisite for the marginal-count argument in T1: the histogram sums over a set of size exactly $q$.

## 5.3. T4: No-Overflow Bounds

**Theorem 5.4 (T4).** For all $q > 0$ and all $x, s_1 < q$ (natural numbers): $1 \leq x + q - s_1$ and $x + q - s_1 < 2q$.

```
theorem no_overflow_bounds {q : ℕ} (hq : q > 0)
    (x s₁ : ℕ) (hx : x < q) (hs : s₁ < q) :
    1 ≤ x + q - s₁ ∧ x + q - s₁ < 2 * q := by
  constructor <;> omega
```

In $\mathbb{Z}_q$, overflow is impossible by construction. T4 bridges to the bit-vector implementation used in actual hardware and in the Z3 encoding: it guarantees that the intermediate value $x + q - s_1$ (the unsigned representation that avoids underflow) fits within the allocated bit-width. Concrete instances for ML-KEM ($2 \times 3{,}329 < 2^{24}$) and ML-DSA ($2 \times 8{,}380{,}417 < 2^{24}$) are discharged by decide.

## 5.4. T5: RNG Bias Characterization

The security guarantee of T1 assumes that the mask $s_1$ is drawn uniformly from $\mathbb{Z}_q$ (assumption A2 in Paper 2). In practice, a $k$-bit random number generator produces values in $\{0, \ldots, 2^k - 1\}$, reduced modulo $q$. When $q \nmid 2^k$, the result is biased.

**Theorem 5.5 (T5: ML-KEM Instance).** For 12-bit RNG ($N = 4096$) reduced modulo $q = 3{,}329$: residue 0 has count 2 and residue 767 has count 1. The bias ratio is exactly 2.

This is verified by native_decide, which compiles the decision procedure to native code and executes it outside the Lean kernel. Unlike decide (which runs inside the kernel), native_decide extends the trusted base to include the native code compiler. The result can alternatively be checked by decide for kernel-only verification at the cost of longer compilation time. We also prove purely kernel-verified universal bounds:

**Theorem 5.6 (T5: Universal Bounds).** For any $N, q > 0$ and residue $r < q$ : $\lfloor N/q \rfloor \leq count(r) \leq \lceil N/q \rceil$. When $q \mid N$, equality holds: $count(r) = N/q$.

The universal bounds are proved by constructing injections between finite sets (Mathlib's Finset.card_le_card_of_injOn).

## 5.5. T6: Universal Non-Tightness Counterexample

Theorem 3.9.1 is sound but not tight: it is possible for a wire to have a constant marginal distribution yet fail to be value-independent. QANARY labels such wires INSECURE_CONSERVATIVE, the analysis correctly errs on the side of caution.

**Theorem 5.7 (T6: Universal Non-Tightness).** For every $q \geq 2$, there exists a wire function $w: \mathbb{Z}_q \times \mathbb{Z}_q \to \{0,1\}$ that has a constant marginal distribution but is not value-independent.

The witness is $w(s_0, s_1) := [s_0 = 0]$ (the indicator function of zero on the first share). The proof has two parts:

1. **Constant marginal.** For any $x, x'$, the sets $\{s_1 : (x - s_1) = 0\}$ and $\{s_1 : (x' - s_1) = 0\}$ have the same cardinality because the translation map $s_1 \mapsto s_1 + (x' - x)$ is a bijection between them (Finset.card_bij).
2. **Not value-independent.** We have $w(0 - 0,0) = [0 = 0] = \text{true}$ but $w(1 - 0,0) = [1 = 0]$, which is false because $1 \neq 0$ in $\mathbb{Z}_q$ when $q \geq 2$. The non-triviality of $\mathbb{Z}_q$ is established by Mathlib's one_ne_zero, which requires only $q \geq 2$.

The Lean 4 formalization of this counterexample follows:

```
theorem converse_fails_universal
    {q : ℕ} [NeZero q] (hq : 2 ≤ q) :
    ∃ (w : WireFunction q Bool),
      HasConstantMarginal w ∧ ¬ValueIndependent w := by
  refine ⟨fun s₀ _ => decide (s₀ = (0 : ZMod q)), ?_, ?_⟩
  · -- HasConstantMarginal: translation bijection
    intro x x' v
    unfold marginalHistogram arithReparam
    apply Finset.card_bij (fun s₁ _ => s₁ + (x' - x))
    · intro s₁ hs₁
      simp only [Finset.mem_filter, Finset.mem_univ,
                 true_and] at hs₁ ⊢
      have : x' - (s₁ + (x' - x)) = x - s₁ := by ring
      rw [this]; exact hs₁
    · intro s₁ _ s₂ _ h; exact add_right_cancel h
    · intro t₁ ht₁
      simp only [Finset.mem_filter, Finset.mem_univ,
                 true_and] at ht₁
```

```
    refine ⟨t₁ + (x - x'),
          Finset.mem_filter.mpr
            ⟨Finset.mem_univ _, ?_⟩, by ring⟩
    have : x - (t₁ + (x - x')) = x' - t₁ := by ring
    rw [this]; exact ht₁
· -- ¬ValueIndependent: one_ne_zero when q ≥ 2
  intro h
  haveI : Fact (1 < q) := ⟨by omega⟩
  have h01 := h (0 : ZMod q) (0 : ZMod q) (1 : ZMod q)
  simp only [arithReparam, sub_zero] at h01
  rw [show decide True = true from rfl,
    eq_comm, decide_eq_true_eq] at h01
  exact absurd h01 one_ne_zero
```

The key technique for universality is replacing Paper 2's decide (which works only at concrete $q$) with Mathlib's one_ne_zero (which works for any $q \geq 2$). The algebraic structure of $\mathbb{Z}/q\mathbb{Z}$ automatically provides non-triviality.

**Remark (Scope of T6).** T6 shows that SADC's conservatism is *inherent*, not an implementation artifact. Any value-independence-based analysis will exhibit this gap for every $q \geq 2$. In particular, the 165 INSECURE_CONSERVATIVE verdicts in Paper 2's ML-KEM Barrett analysis are not artifacts of QANARY's implementation, Theorem 5.7 proves that this gap is unavoidable. The conservative direction is the safe one for a security tool: false positives (labeling a secure wire as potentially insecure) are acceptable; false negatives are not.

# 6. Implications

## 6.1. For FIPS 140-3 Certification

FIPS 140-3 [22] certification of cryptographic modules requires evidence that side-channel countermeasures are effective. A universal proof eliminates the need to re-verify for each parameter set: a single kernel-verified proof covers ML-KEM ($q = 3{,}329$), ML-DSA ($q = 8{,}380{,}417$), and any future NIST PQC standard over any prime field. This directly reduces evaluation burden for Cryptographic Module Validation Program (CMVP) testing laboratories. Concretely, the 165 wires labeled INSECURE_CONSERVATIVE in Paper 2's ML-KEM Barrett reduction analysis carry a soundness guarantee that is now universally verified: if SADC labels a wire secure, the security conclusion holds for $q = 3{,}329$ not merely because we checked it at $q = 5$, but because Theorem 4.1 is valid for every positive modulus.

## 6.2. For Design Portability

Post-standardization parameter updates, changing $q$ in response to cryptanalytic advances, or adopting new lattice-based schemes, do not invalidate the proof. The universal quantification over $q$ means that QANARY's soundness guarantee transfers automatically.

## 6.3. For Tool Trust

The shift from "trust Z3 and CVC5 on $2^{25}$ instances" to "trust the Lean 4 kernel on a universally quantified statement" is qualitatively different for high-assurance applications. For the core soundness theorems (T1–T4, T5 universal bounds, T6), the Lean 4 kernel, a small, formally specified C++ core designed for auditability, is the sole element in the trusted computing base. T5's ML-KEM instance additionally depends on the native code compiler. By contrast, Z3 and CVC5 are complex systems (hundreds of thousands of lines each) whose correctness is not formally verified. Table 3 contrasts the two approaches' scope, trusted base, and proof artifacts.

**Table 3.** Build and verification comparison.

| Approach | Scope | Trusted Base | Proof Artifact |
|---|---|---|---|
| Z3/CVC5 (Paper 2) | $q = 5$, $2^{25}$ wires | Z3, CVC5, Python encoding | 363 dual-solver queries (Z3+CVC5), 0 disagreements |
| Lean 4 (this work) | All $q > 0$, all $w$, all $\beta$ | Lean 4 kernel (T1–T4, T5 bounds, T6); native compiler (T5 instance) | 1,739 build tasks, 0 sorry |

## 6.4. For the Field

The five-line proof in Section 4.1 is evidence for a broader claim: ring theory is the natural abstraction layer for arithmetic masking verification.

The Z3/CVC5 encoding required bit-vector arithmetic, explicit overflow guards, and URem operations because SMT solvers operate on concrete representations. Lean 4's $\mathbb{Z}_q$ is the quotient ring $\mathbb{Z}/q\mathbb{Z}$, where modular arithmetic is definitional. The reparametrization $s_0 = x - s_1$ is not a computational procedure to be checked for overflow, it is a ring identity. The proof that reparametrization is well-defined (T3) is one call to sub_add_cancel, a CommRing axiom.

This observation suggests that other properties of arithmetic masking, composition theorems, higher-order masking, mixed Boolean-arithmetic conversions, may admit similarly concise proofs when formulated in the appropriate algebraic setting.

We state one such conjecture precisely. An NTT butterfly stage takes two masked inputs $(a, a')$ and $(b, b')$ and produces two masked outputs via $c = a + tb \pmod q$, $d = a - tb \pmod q$ for twiddle factor $t$. If each stage independently satisfies value-independence (Theorem 4.1), does the composition of $\log_2 n$ stages in a full n-point NTT preserve first-order probing security? We conjecture that it does, and that the proof follows from the ring homomorphism structure of NTT stages: each stage is a $\mathbb{Z}_q$-linear map on shares, and the composition of $\mathbb{Z}_q$-linear maps is $\mathbb{Z}_q$-linear. If this conjecture holds, the five-line proof structure of Theorem 4.1 would extend to a machine-checked composition theorem for full NTT pipelines, a result with no precedent in the masking verification literature.

## 7. Limitations and Future Work

The two primary limitations below are related: together they constitute the single remaining gap between the formalized result (constant marginal histogram) and a complete machine-checked proof of $I(x; w) = 0$ under fresh randomness. Paper 4 will close this gap.

**(i) $r$-free sub-theorem.** T1 proves the $r$-free version of Theorem 3.9.1: it marginalizes only over the mask $s_1$, not over fresh randomness $r$. The extension to the $r$-bearing version follows by the law of total probability: if the $r$-free marginal is constant in $x$ for every fixed $r$, then marginalizing over $r$ preserves the property. This bridging argument is a standard one-paragraph derivation; we defer its formalization to future work (ProbBridge.lean). Importantly, the Z3/CVC5 proofs in Paper 2 were also $r$-free in the same sense, this limitation is not introduced by the Lean formalization but is inherited from SADC's structural semantics, which evaluates combinational logic for fixed randomness.

**(ii) Information-theoretic formalization.** Observation 4.2 ( $I(x; w) = 0$ ) follows immediately from the constant histogram, but Mathlib currently lacks definitions of Shannon entropy and mutual information for finite types. We prove the histogram-level statement, which is equivalent. Formalizing $I(x; w) = 0$ directly would require either contributing entropy definitions to Mathlib or developing them within the project.

**(iii) RC toolchain.** The proof suite was developed on Lean 4.30.0-rc1. We intend to migrate to a stable Lean 4 release before venue submission.

**(iv) Higher-order masking.** The current formalization addresses first-order masking ($d = 1$, two shares). Extension to $d$-th order masking ($d + 1$ shares) requires a $d$-dimensional version of the reparametrization and a corresponding multi-marginal argument.

**(v) Glitch-extended probing model.** The standard probing model assumes the adversary observes one wire per clock cycle. The glitch-extended model (relevant for hardware implementations) allows observation of multiple wires in combinational logic cones. Extending the universal proof to this model is an open problem.

## 8. Conclusion

We have presented, to our knowledge, the first machine-checked universal proof that value-independence implies distributional security ($r$-free sub-theorem) for arithmetic masking in NTT-based PQC hardware. The proof, mechanized in Lean 4 with Mathlib, closes the finite-domain gap explicitly identified in the QANARY framework: a soundness result that was previously verified only at $q = 5$ by exhaustive SMT enumeration now holds for every positive modulus.

The proof suite comprises nine theorems across eight IDs (T1, T1', T2, T3, T3', T4, T5, T6), all sorry-free: 1,739 build jobs, zero sorry, zero errors. The supporting theorems establish reparametrization correctness, bijectivity, overflow bounds, RNG bias characterization, and a universal counterexample showing that the converse of Theorem 3.9.1 fails for every $q \geq 2$.

The result that required 33,554,432 Boolean evaluations in Z3 required five lines in Lean 4. This is not because the problem became easier. It is because we found the language in which the result is natural: the commutative ring axioms of $\mathbb{Z}/q\mathbb{Z}$.

Paper 4 in this series will formalize the probabilistic bridge from the $r$-free sub-theorem established here to the full $r$-bearing distributional security guarantee, completing the formal foundations for QANARY's verification hierarchy.

## 9. Code and Data Availability

The complete Lean 4 proof suite supporting this paper is publicly available at https://github.com/rayiskander2406/qanary-universal-masking-proofs-arXiv-2604.18717 under the MIT license, with an archival snapshot of version v1.0.0 deposited on Zenodo at https://doi.org/10.5281/zenodo.19689480. The repository contains the complete sorry-free mechanization of all nine theorems (T1, T1', T2, T3, T3', T4, T5, T5 universal bounds, T6) reported in this paper, with lean-toolchain pinned to leanprover/lean4:v4.30.0-rc1 and lakefile.lean pinning Mathlib to commit 322515540d7f. A single lake build invocation regenerates the headline result (1,739 build tasks, 0 sorry, 0 errors) on a clean checkout; no other dependencies are required. We intend to submit the artifact for venue artifact evaluation upon acceptance.

The companion papers in this series are archived at https://doi.org/10.5281/zenodo.19625392 (Paper 1, structural dependency analysis; reference [1]) and https://doi.org/10.5281/zenodo.19508454 (Paper 2, partial NTT masking security margin analysis; reference [2]).